\documentclass[a4paper,11pt]{article}

\usepackage{jcappub} 

\usepackage[T1]{fontenc} 
\usepackage{ae,aecompl,bm}
\usepackage{xcolor}

\newcommand{\OHI}{\Omega_{\rm H_{\rm I}}}
\newcommand{\HI}{{\rm H}_{\textsc{I}}}

\title{Detectability of CMB Weak Lensing and HI Cross Correlation and constraints on cosmological parameters}

\author[1]{Shoichiro Tanaka,}
\author[2]{Shintaro Yoshiura,}
\author[1]{Kenji Kubota,}
\author[1,3]{Keitaro Takahashi,}
\author[4,5]{Atushi J. Nishizawa,}
\author[4,6]{Naoshi Sugiyama}

\affiliation[1]{Graduate School of Science and Technology, Kumamoto University, Kumamoto, 860-8555, Japan}
\affiliation[2]{The University of Melbourne, School of Physics, Parkville, VIC 3010, Australia}
\affiliation[3]{International Research Organization for Advanced Science and Technology, Kumamoto University, Kumamoto, 860-8555, Japan}
\affiliation[4]{Department of Physics, Nagoya University, Aichi, 464-8602, Japan}
\affiliation[5]{Institute for Advanced Research, Nagoya University, Aichi, 464-8602, Japan}
\affiliation[6]{Kavli Institute for the Physics and Mathematics of the Universe, The University of Tokyo, Chiba, 277-8583, Japan}
\emailAdd{206d7151@st.kumamoto-u.ac.jp}

\abstract{Neutral hydrogen (H$_{\rm I}$) intensity mapping is capable of measuring redshift evolution of H$_{\rm I}$ density parameter $\Omega_{\rm H_{\rm I}}$, which is an important parameter to understand structure formation in the post-reionization epoch. Future H$_{\rm I}$ observation with Square Kilometre Array (SKA) can significantly improve constraints on the parameter. However, the observation of H$_{\rm I}$ suffers from the contamination from extremely bright foreground emissions, and it is necessary to consider a signal validation method complementary to the measurement of H$_{\rm I}$ auto power spectrum. In this work, we propose to take a cross correlation between a 21cm-line intensity map and a convergence map reconstructed from observation of the cosmic microwave background by Planck and estimate expected ideal constraints on cosmological parametersby ignoring the foreground contamination. We find that the SKA1-mid operated in single-dish mode has a sufficient capability to detect the cross correlation on the large scales. Further, by Fisher analysis assuming a constant linear bias of H$_{\rm I}$, $b$, and SKA1-mid observation of 1,000 hours for each of Band 1 and 2, we show that $\Omega_{\rm H_{\rm I}} b$ can be constrained with a precision of $6-13\%$ at a wide range of redshifts of $0.0<z<2.0$, if we fix other cosmological parameters (density parameter of cold dark matter, $\Omega_{\rm c} h^{2}$, spectral index of primordial fluctuations, $n_s$, and Hubble constant, $H_0$) using values from the Planck observations. On the other hand, small-scale measurements with interferometer mode of SKA1-mid will not have a significant impact on constraining the parameters with 1,000 hours of observation time for each of Band 1 and 2, due to the limited resolution of the CMB lensing of Planck.}

\begin{document}
\maketitle
\flushbottom

\section{Introduction}
\label{sec:intro}

Recent remarkable progress of cosmological observations has tightly constrained cosmological parameters. In order to further improve the precision, one direction to go is to measure the dark matter distribution precisely across a large cosmological volume. In the post-reionization epoch ($z$<6), almost all of the hydrogen in the intergalactic medium (IGM) is ionized, and small amount of neutral gas remain within a high density region, typically inside galaxies \citep{zwaan,zwaan1,briggs}. Therefore, the 21 cm line from neutral hydrogen (H$_{\rm I}$) can trace the dark matter distribution. In fact, the measurement of the 21 cm signal is one of the key sciences of upcoming radio telescopes such as the Square Kilometer Array (SKA) \citep{santos}. The high angular-resolution and wide frequency coverage of the SKA allow us to measure the three dimensional distribution of neutral hydrogen across a wide range of redshifts.

The 21\,cm signal from highly distant galaxies is faint, and the direct detection has not been achieved yet for high redshift. Furthermore, foreground emission from our Galaxy contaminates the cosmological 21\,cm signal and therefore the foreground removal is a serious challenge to be solved \citep{wolz}. Alternatively, the cross correlation of 21cm line with other cosmological probes can avoid the foreground contamination. For example, Chang et al.(2010)\citep{chang} has measured the cross correlation power spectrum between the cumulative 21cm line from the Green Bank Telescope and the distribution of 10,000 galaxies at $z<1$ which are measured using DEEP2 optical galaxy redshift survey \citep{marinoni}. This has obtained a positive correlation and showed the statistical detection of 21 cm signal at the 4 $\sigma$ level.

The amount of H$_{\rm I}$ mass density $\Omega_{\rm H_{\rm I}}$ is an important quantity in order to understand the property and evolution of galaxies at the post-reionization epoch. For example, spectrum of high-$z$ quasars are absorbed by damped Lyman-$\alpha$ system, and the absorption gives insights on the amount of H$_{\rm I}$. Since the quasars are measured over a wide range of redshift, $\Omega_{\rm H_{\rm I}}$ has been constrained at various redshifts. Moreover, the cross correlation between 21\,cm line and galaxy has constrained $\Omega_{\rm H_{\rm I}}$ at lower redshift ($z<3.5$) \citep{chang,Padmanabhan}. However, the constraints on $\Omega_{\rm H_{\rm I}}$ are not strong compared to standard cosmological parameters.

The cross correlation between the convergence field of cosmic microwave background (CMB) lensing and 21\,cm signal has been first formulated by Guha Sarkar(2010)\citep{guha}. The CMB photon is deflected by the gravitational potential of large-scale structure and secondary CMB anisotropies are induced. The CMB lensing is sensitive to the structure at $z<3$ and thus well correlated with the 21\,cm signal from post-reionization epoch. The convergence field of CMB can be reconstructed from temperature fluctuation with high angular resolution. Furthermore, the B-mode polarization fluctuation can also be used as an alternative way to reconstruct the convergence field. Recent CMB experiments tend to have a better sensitivity to small scale fluctuation and polarization, which can be suitable for cross correlation study with the 21\,cm line.

Guha Sarkar(2010)\citep{guha} has found that the 21\,cm-CMB lensing cross power spectrum (CPS) will be a promising measure for detecting the 21\,cm signals; however, the thermal noise, which in practice dominates the error budget after securely removing the foreground emission, has not been considered in the forecast.

In this work, we study the detectability of the 21\,cm-CMB lensing CPS assuming a SKA observation and the convergence map measured by Planck. Furthermore, we perform the Fisher analysis and estimate the expected constraints on cosmological parameters, especially focusing on $\Omega_{\rm H_{\rm I}} b$, where $b$ is the bias between H$_{\rm I}$ and cold dark matter density fluctuations.

This paper is organized as follows. Section \ref{sec:method} details the 21\,cm-CMB lensing cross correlation, and briefly describes the error formulae and Fisher analysis. In Section \ref{sec:result}, we give the main results on the detectability of the CPS and forecasts for parameter constraints. Finally, Section \ref{sec:summary} is devoted to the summary. Throughout the paper, we assume $\Lambda$CDM model and adopt cosmological parameters consistent with \citep{planck18} unless otherwise stated, $(\Omega_{b}h^{2}, \Omega_{\rm c}h^{2}, n_{s}, H_{0}) = (0.0224, 0.119, 0.967, 67.7)$. Furthermore, the fiducial value of $\Omega_{\rm H_{\rm I}} b$ is set to $2.0\times10^{-3}$.

\section{21 cm-CMB lensing Cross Correlation}
\label{sec:method}
\subsection{auto and cross power spectra}
The CMB lensing is quantified by the convergence field $\kappa$ and can be written as, 
\begin{eqnarray}
\kappa(\hat{\bm {n}}) &=& \frac{3}{2} \Omega_{m0} \left(\frac{H_0}{c}\right)^{2} \int_{0}^{x_{\rm LSS}} d\chi \, F(\chi) \delta (x{\hat{\bm n}}),\\
F(\chi) &=& \frac{S_{\rm K}(\chi_{\rm LSS}-\chi)S_{\rm K}(\chi)D_{+}(\chi)}{S_{\rm K}(\chi_{\rm LSS})a(\chi)},
\end{eqnarray}
where $\Omega_{m0}$ is the matter density parameter, $H_0$ is the Hubble constant, $z_{\rm LSS}$ is the redshift at the last scattering surface, $a(z)$ is the scale factor, $\chi$ is comoving angular diameter distance, $\delta$ is the matter fluctuation at the present day and $D_+$ is the growth factor. 

Since the convergence field is calculated as an integral of density fluctuation from the last scattering surface up to the present, we can not measure the density fluctuations at a particular redshift. However, the cross correlation between the CMB lensing and the H$_{\rm I}$ can extract the effect of CMB lensing at given redshifts. 

Following Guha Sarkar(2010)\citep{guha}, here we briefly revisit the formulation of 21\,cm-CMB lensing CPS. Using spherical harmonics, the convergence field contributed from the large-scale structure can be expanded as,
\begin{equation}
\kappa(\hat{\bm{n}}) = \sum_{\ell,m}^{\infty} a_{\ell m}^{\kappa}Y_{\ell m}(\hat{\bm{n}}),
\end{equation}
where $Y_{\ell m}(\hat{\bm{n}})$ is the spherical harmonic function. 
Then, the coefficients $a_{\ell m}^{\kappa}$ is obtained as,
\begin{equation}
a_{\ell m}^{\kappa} = \int d\omega_{(\hat{\bm{n}})}\kappa(\hat{\bm{n}})Y_{\ell m}^{\ast}(\hat{\bm{n}}),
\end{equation}
where $\omega_{\hat{\bm{n}}}$ is the solid angle. The coefficients can be described using Raleigh expansion
as
\begin{eqnarray}
a_{\ell m}^{\kappa}&=&6\pi \Omega_{\rm m0}\left(\frac{H_{0}}{c}\right)^2(-i)^{\ell}\int \frac{d^{3}\bm{k}}{(2\pi)^{3}} \nonumber\\
&\times&\int_{0}^{\chi_{\rm LSS}}d\chi F(\chi)\delta(\bm{k})j_{\ell}(kr)Y_{\ell m}^{\ast}(\hat{\bm{k}}),
\label{eq:kappa1}
\end{eqnarray}
where $\delta(\bm{k})$ is the Fourier transform of $\delta(\bm{r})$, and $j_{\ell}(\chi)$ is the spherical Bessel function.

For the 21\,cm observation, we measure the brightness temperature, $T(z)$, which can be expressed as
\begin{eqnarray}
    T(\nu , \hat{\bm n}) &=& \bar{T}[1+\delta_{\rm H_{\rm I}}(z, \hat{\bm n}r_{\rm H_{\rm I}})]\left(1-\frac{T_{\gamma}}{T_{S}}\right)\nonumber\\
    &\times&\biggl[1-\frac{1+z}{H(z)}\frac{\partial v(z, \hat{\bm n}r_{\rm H_{\rm I}})}{\partial r}\biggl],
\end{eqnarray}
where $T_{\gamma}$ and $T_{S}$ are the CMB temperature and the spin temperature, respectively. The mean brightness temperature and mean neutral hydrogen fraction are scaled as

\begin{equation}
\bar{T}(z) = 4.0~{\rm mK}~(1+z)^{2}\bar{x}_{\rm H_{\rm I}}(z)\left(\frac{\Omega_{b0}h^{2}}{0.02}\right)\left(\frac{70}{H_0 [\rm km/s/Mpc]}\right),
\end{equation}
and
\begin{equation}
\bar{x}_{\rm H_{\rm I}} = 50 \Omega_{\rm H_{\rm I}} h^{2} \left(\frac{0.02}{\Omega_{b}h^2}\right),
\end{equation}
respectively.
Now, we see that the brightness temperature fluctuation can be generated both from density fluctuation and velocity gradient of the neutral hydrogen clouds. We assume that the spatial variation of the spin temperature is negligible at our redshift ranges and that it is significantly higher than the CMB temperature, $T_S \gg T_{\gamma}$, which are valid at low redshifts considered here ($z \leq 3$).

Working in the Fourier space makes things much simpler, i.e.,
\begin{eqnarray}
    \frac{\partial v(z, \hat{\bm n}r_{\rm H_{\rm I}})}{\partial r}
    &\rightarrow& -\frac{k_{\|}^2}{k^2}\dot{D_+}(z)\delta(\bm k) \nonumber \\
    &=&-fH(z)\mu^{2}\delta(z,\bm k),
    \label{eq:vg}
\end{eqnarray}
where $f\equiv{\rm d}\!\ln D_+/{\rm d}\!\ln a = \dot{D_+}(z)/(D_{+}(z)H(z)) \simeq \Omega_m^{0.6}$, $\mu$ stands for the cosine of angle between the line-of-sight direction $\hat{\bm n}$ and the wave vector ($\mu = \hat{\bm k} \cdot \hat{\bm n}$).
With an assumption of constant bias of $\HI$, $\delta_{\HI} = b \delta$, and no velocity bias, 
we obtain
\begin{equation}
a_{\ell m}^{\rm H_{\rm I}} = 4\pi \bar{T}(z)(-i)^{\ell} \int \frac{d^{3}\bm{k}}{(2\pi)^{3}}\delta(\bm{k},z)J_{\ell}(kr)Y_{\ell m}^{\ast}(\hat{\bm{k}}),
\label{eq:21--cm}
\end{equation}
where $J_{\ell}$ is defined as,
\begin{equation}
J_{\ell}(\chi) \equiv \left(b - f\frac{d^{2}}{d\chi^{2}}\right)j_{\ell}(\chi).
\end{equation}

Using Eq.~(\ref{eq:kappa1}), Eq.~(\ref{eq:21--cm}) and the Limber approximation for large $\it {\ell}$ in Fourier space, we can describe the CPS as,
\begin{eqnarray}
\label{eq:CPS}
C_{\ell}^{\rm H_{\rm I}-\kappa} &\approx& \frac{\pi}{2}A(z_{\rm H_{\rm I}})b\frac{F(\chi(z_{\rm H_{\rm I}}))}{\chi^{2}(z_{\rm H_{\rm I}})}P\left(\frac{\ell}{\rm r_{\rm H_{\rm I}}}\right),\\
A(z) &=& \frac{3}{\pi} \Omega_{\rm m0} \left(\frac{H_0}{c}\right)^{2} \bar{T}(z)D_{+}(z),
\end{eqnarray}
where $r_{\rm H_{\rm I}}$ is the comoving distance at a redshift $z_{\rm H_{\rm I}}$ and $P(k)$ is the present-day dark-matter linear power spectrum.
We use a public code \texttt{CAMB} \citep{camb, camb2} for the theoretical prediction of $P(k)$. We assume a constant linear bias, $b=2$, at all redshifts. Since the CMB lensing has a broad kernel along the line of sight, modes parallel to the line of sight are cancelled out and thus we can always set $\mu=0$.

For convenience, we describe the auto power spectra for CMB lensing and 21\,cm fluctuation. The convergence power spectrum for large $\it{\ell}$ is approximately given by,
\begin{equation}
C_{\ell}^{\kappa}\approx \frac{9}{4}\Omega^{2}_{\rm m0}\left(\frac{H_0}{c}\right)^{4}\int d\chi \frac{F^{2}(\chi(z))}{\chi^{2}(z)}P\left(\frac{\ell}{\chi(z)}\right).
\label{eq:kappa}
\end{equation}
The angular power spectrum of 21\,cm signal, $C_{\ell}^{\rm H_{\rm I}}(z_{\rm H_{\rm I}})$, is a direct observational estimator of the H$_{\rm I}$ fluctuation at a redshift $z_{\rm H_{\rm I}}$. The $C_{\ell}^{\rm H_{\rm I}} (z_{\rm H_{\rm I}})$ with the flat sky approximation \citep{datta}, valid for $\it {\ell} \gtrsim 8$, is given by, 
\begin{equation}
C_{\ell}^{\rm H_{\rm I}}(z_{\rm H_{\rm I}})=\frac{\bar{T}^2}{\pi r^{2}_{\rm H_{\rm I}}}D^{2}_{+}\int_{0}^{\infty}dk_{\|}[b+f\mu^{2}]^{2}P(k),
\label{eq:21--cmAPS}
\end{equation}
where the wave-number vector ${\bm k}$ has magnitude $k=\sqrt{k_{\|}^{2}+l^{2}/r_{\rm H_{\rm I}}^{2}}$, where $k_\|$ is the wave number along the line of sight. 

For 21cm observation, the telescope's finite size causes the instrumental noise and beam smoothing. These erase all modes below the telescope resolution, which damps small-scale power of the 21cm-line auto-power spectrum and 21cm-CMB lensing cross-power spectrum. For an angular Gaussian beam, the harmonic coeffcients of the beam can be written as, \citep{Witzemann}
\begin{equation}
B_{\ell}=\exp \left(-\frac{\ell(\ell+1) \theta_{\mathrm{B}}^{2}}{16 \log 2}\right),
\label{eq:beam}
\end{equation}
where $\theta_{\mathrm{B}}$ is the solid angle of the primary beam. As a result, for SKA-mid single dish mode,  Eqs.~(\ref{eq:CPS}) and (\ref{eq:21--cmAPS}) are modified as,
\begin{equation}
C_{\ell}^{\rm H_{\rm I}-\kappa} \approx
\frac{\pi}{2}A(z_{\rm H_{\rm I}})b B_{\ell} \frac{F(\chi(z_{\rm H_{\rm I}}))}{\chi^{2}(z_{\rm H_{\rm I}})} P\left(\frac{\ell}{\rm r_{\rm H_{\rm I}}}\right),
\label{eq:CPS_2}
\end{equation}
\begin{equation}
C_{\ell}^{\rm H_{\rm I}}(z_{\rm H_{\rm I}}) =
\frac{\bar{T}^2}{\pi r^{2}_{\rm H_{\rm I}}} D^{2}_{+} B^2_{\ell} \int_{0}^{\infty}dk_{\|}[b+f\mu^{2}]^{2}P(k),
\label{eq:21--cmAPS_2}
\end{equation}
\subsection{error formula}
In order to argue the detectability of the 21\,cm-CMB lensing CPS, we use the following error formula on the CPS \citep{guha,guha16},
\begin{equation}
(\Delta C_{\ell}^{{\rm H_{\rm I}}-\kappa})^2
= \frac{(C_{\ell}^{\kappa}+N_{\ell}^{\kappa})(C_{\ell}^{\rm H_{\rm I}}+N_{\ell}^{\rm H_{\rm I}})+(C_{\ell}^{\rm H_{\rm I}-\kappa})^{2}}{(2 \ell + 1) N_{c} f_{\rm sky} \Delta \ell},
\label{eq:error}
\end{equation}
where $\Delta \ell$ is the width of binned multipole $\ell$, $f_{\rm sky}$ is the sky fraction of the survey field, $N_c$ is the number of channels for the 21\,cm-line observation. The second term of the Eq.~(\ref{eq:error}) is usually smaller than the first term.

We use the minimum variance reconstruction approximate noise power spectrum \footnote{Noise power spectrum can be found in\\ https://wiki.cosmos.esa.int/planck-legacy-archive/index.php/Lensing}  
\citep{planck18} to evaluate the error of the convergence map. The Planck survey has achieved a reconstruction of the convergence filed over 70 \% of the sky. 

A next-generation radio telescope SKA is planned to be constructed in the 2020s, and the low-redshift 21\,cm signal ($z < 3$) will be measured by the SKA-mid. The SKA1-mid, the first phase of SKA-mid, in South Africa consists of 190 new dishes and MeerKAT 64 dishes and will be operated as single-dish mode (SD) and interferometer mode (IF). The SD mode covers a large fraction of the sky and can be used to perform large-scale intensity mapping ($\ell \lesssim 250$), while the IF mode realizes high angular resolution suitable for measuring small-scale power spectrum ($\ell \gtrsim 300$).

For the SD-mode observation, the noise term is written as \citep{bull},
\begin{eqnarray}
N^{\rm H_{\rm I}}_{\ell,\rm SD} = \frac{\lambda^4\, T^2_{\rm sys} }{A^2_{\rm eff}\, \Delta\nu \,t_{\rm tot}} \frac{S_{\rm sky}}{\theta_{\rm B}^4},
\label{eq:noiseska1}
\end{eqnarray}
where $\lambda$ is the observed wavelength, $T_{\rm sys}$ is the system temperature, $A_{\rm eff}$ is the effective collecting area of a dish, $S_{\rm sky}$ is the survey area, $\theta_{\rm B}$ is the solid angle of primary beam and $t_{\rm tot}$ is the total observation time of the survey. 

For the IF-mode observation, we assume a observation of a single pointing, and the noise is given as,

\begin{eqnarray}
N^{\rm H_{\rm I}}_{\ell,\rm IF} = \frac{\lambda^4\, T^2_{\rm sys} }{A^2_{\rm eff}\, \Delta\nu\, t_{\rm \rm obs}\, n(|{\bf u}|,\nu)},
\label{eq:noiseska2}
\end{eqnarray}

where $n(|{\bf u}|,\nu)$ is the number density of baseline contributing to $l = 2 \pi |{\bf u}|$ and $t_{\rm obs}$ is the observation time per pointing. For the baseline distribution, we refer the SKA1 baseline design \footnote{Braun, R. et al. (2017). Tech. rep. SKA-TEL-SKO-0000818-01. SKA Organization.}. Based on the baseline design, we assume the following array distribution: 76, 26, 31 and 21 antennas are uniformly distributed in radius $R <$ 400 m, 400 m $< R <$ 1,000 m, 1,000 m $< R <$ 2,500 m and 2,500 m $< R <$ 4,000 m, respectively.

\begin{table*}
\caption{SKA1-mid survey parameters for each redshift bin: redshift range, corresponding frequency range, number of frequency channels, beam size, effective collecting area of a dish divided by system temperature and multipole coverage by SD and IF modes. Other common parameters are, channel width $\Delta \nu = 10~{\rm MHz}$, sky fraction and survey area of SD-mode surveys $f_{\rm sky}^{\rm (SD)} = 0.6$ and $S_{\rm sky} = 7.61~{\rm sr}$, respectively. Note that the fiducial survey area of IF-mode observation is equal to the beam size.}
\scalebox{0.7}{
 \begin{tabular}[width=\columnwidth]{|c|c|c|c|c|c|c|} \hline
 redshift & frequency [MHz] & $N_{\rm c}$ & beam size $\theta_{\rm B}$ [rad] & $A_{\rm eff}$/$T_{\rm sys}$ [${\rm m}^2/{\rm K}$] & multipole (SD) & multipole (IF) \\ \hline \hline
  0.5 ($0.0 < z < 1.0$) & 1,420 - 710 & 71.0 & 0.021 & 6.04 & 8$\leq \ell \leq$598 & 599$\leq \ell \leq$2000  \\
  1.5 ($1.0 < z < 2.0$) & 710 - 473 & 23.7 & 0.035 & 4.61 & 8$\leq \ell \leq$358 & 359$\leq \ell \leq$2000  \\
  2.5 ($2.0 < z < 3.0$) & 473 - 355 & 11.8 & 0.049 & 3.12 & 8$\leq \ell \leq$256 & 257$\leq \ell \leq$2000  \\  \hline 
\end{tabular}
}
\label{table:survey}
\end{table*}

We consider a redshift range of $0 < z < 3$ and divide it into three bins: $0 < z < 1$, $1 < z < 2$ and $2 < z < 3$. The corresponding frequency ranges are 1,420 MHz - 710 MHz, 710 MHz - 473 MHz and 473 MHz - 355 MHz, respectively. The higher-frequency half of the first frequency range ($0 < z < 1$) can be observed with SKA1-mid Band 2, which covers from 950 MHz to 1,760 MHz. On the other hand, the other frequency ranges can be observed with SKA1-mid Band 1, which covers from 350 MHz to 1,050 MHz. We assume 1,000 hours of observation for each mode (SD and IF) and Band (1 and 2) so that the total observation time is 4,000 hours to collect the data at all three redshift bins.

In this work, we assume the channel width of $\Delta \nu = 10~{\rm MHz}$ to optimize the detectability. The number of channels, $N_{\rm c}$, for each redshift bin can be calculated as $B/\Delta \nu$ where $B$ is the corresponding frequency range. Since the 21\,cm-line signal is considered to be statistically independent at different frequencies, the error is reduced by a factor of $N_{c}$.

Concerning the sky fraction of surveys, we assume $f_{\rm sky}^{\rm (SD)} = 0.6$ for SD mode, which gives $S_{\rm sky} = 7.61~{\rm sr}$. The multipole range which can be probed by SD mode is determined by the survey area and the size of the primary beam. In fact, the beam size depends on the frequency and the multipole ranges covered by SD mode are $8 < \ell < 598$, $8 < \ell < 358$ and $8 < \ell < 256$ for $0 < z < 1$, $1 < z < 2$ and $2 < z < 3$, respectively.

On the other hand, larger multipoles (smaller scales) are covered by IF-mode observations. For IF mode, we assume that all the observation time is devoted to a single field-of-view so that the survey area is equal to the beam size. We take multipoles up to 2,000 into account because the error on convergence map are so large at even larger multipoles (Fig.\ref{fig:APSK}) and these modes do not give useful information as shown below . The survey parameters are summarized in Table \ref{table:survey}.

The signal-to-noise (S/N) ratios for different multipoles can be combined to give the total S/N ratio as,
\begin{eqnarray}
\left(\frac{S}{N}\right)^{2}
&\equiv& \sum_{\ell} \left(\frac{C_{\ell}^{\rm H_{\rm I}-\kappa}}{\Delta C_{\ell}^{\rm H_{\rm I}-\kappa}}\right)^{2} \nonumber \\
&=& \sum_{\ell} \frac{(2 \ell + 1) N_{c} f_{\rm sky} (C_{\ell}^{\rm H_{\rm I}-\kappa})^2}{(C_{\ell}^{\kappa}+N_{\ell}^{\kappa}) (C_{\ell}^{\rm H_{\rm I}}+N_{\ell}^{\rm H_{\rm I}})+(C_{\ell}^{\rm H_{\rm I}-\kappa})^{2}}.
\label{eq:SN2}
\end{eqnarray}
As noted before, the redshift bin of $z=0-1$ is covered by SKA1-mid Band 1 and 2, and its S/N ratio can be obtained by combining those of Band 1 and 2 in the same way as in the above equation.

While foregrounds do not contribute to the average of the measurement of the CPS and it's error is reduced only for the independent mode because they do not correlate statistically with the convergence map, they do induce the variance of the measurement and contribute to the error. The foreground can be removed as it is spectrally smooth, but the perfect removal is impossible and residuals would still remain to some extent \citep{wolz}. Thus, the foreground removal is one of the most important and challenging issue for the 21\,cm data analysis. In this paper, for the purpose of simplicity, we ignore the contribution of residuals to the error budget.
\subsection{Fisher analysis}
We perform the Fisher analysis to estimate expected constraints on cosmological parameters. In this work, we focus on $\Omega_{\rm H_{\rm I}}$ and three cosmological parameters ($\Omega_{\rm c}h^{2}, n_{s}$ and $H_0$). In fact, due to the bias between the cold dark matter and H$_{\rm I}$ gas fluctuations, $\Omega_{\rm H_{\rm I}}$ always appears as the combination $\Omega_{\rm 
H_{\rm I}} b$ in the observable quantities of cross correlation and $\Omega_{\rm H_{\rm I}}$ cannot be constrained separately. Therefore, we treat $\Omega_{\rm H_{\rm I}} b$ as a single parameter. If the bias $b$ is constrained by other methods, a direct constraint on $\Omega_{\rm H_{\rm I}}$ would be obtained by combining it with the constraint from the cross correlation discussed here.

The Fisher matrix on parameter $p_i$ and $p_j$ is calculated as \citep{tegmark,dan},
\begin{equation}
F_{ij}=\frac{1}{2}\Biggl\langle\frac{\partial^2 
\ln {\cal L}
}{\partial p_i \partial p_j} \Biggl\rangle \Biggl |_{\bm{p} = \bm{p}_{\rm fid}} ,
\end{equation}
where the log likelihood
is given by,
\begin{equation}
\ln {\cal L}(\bm{p})
= \sum_{\ell = \ell_{\rm min}}^{\ell_{\rm max}}
\left[
\frac{C_{\ell}^{\rm H_{\rm I}-\kappa}(\bm {p})-C_{\ell}^{\rm H_{\rm I}-\kappa}(\bm {p}_{\rm fid})}
{\Delta C_{\ell}^{{\rm H_{\rm I}}-\kappa}(\bm {p}_{\rm fid})}
\right]^2.
\end{equation}
Here, the parameter vector is given as ${\bm{p}} = (\Omega_{\rm H_{\rm I}} b,\Omega_{\rm c}h^{2},n_s, H_0)$ and ${\bm{p}}_{\rm fid}$ represents the fiducial parameter set. The error on the parameters are evaluated using covariance matrix, which corresponds to the inverse matrix of Fisher matrix,
\begin{equation}
\label{eq:fihser_inverse}
    \sigma(p_i) = 
    \sqrt{[\bm{F}^{-1}]_{ii}}.
\end{equation}

Finally, note that $\sigma_8$ completely degenerates with $\Omega_{\rm H_{\rm I}} b$ in the cross power spectrum. Therefore we assume $\sigma_8$ is well constrained by other observations and we do not consider it as a free parameter in our Fisher analysis. Further, we note that the three cosmological parameters ($\Omega_{\rm c}h^{2}, n_{s}$ and $H_0$) have also been constrained much better compared with $\Omega_{\rm H_{\rm I}}$ \citep{planck18}.
\section{Results \& Discussion}
\label{sec:result}
\begin{figure}
\includegraphics[width=\columnwidth]{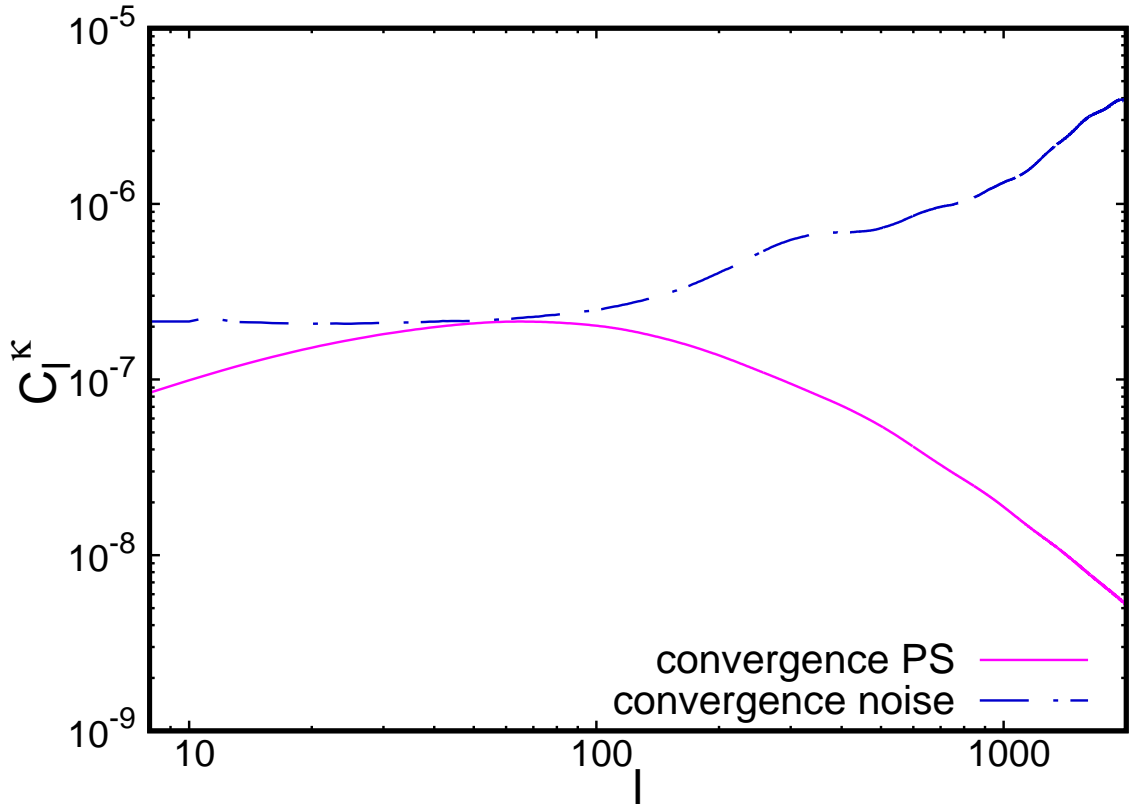}
\caption {Convergence power spectrum $C_{\ell}^{\kappa}$ calculated from Eq.~(\ref{eq:kappa}) (solid) and the observation noise of Planck (dot-dashed) \citep{planck18}.}
\label{fig:APSK}
\end{figure}

In this section, we present our main results. Fig.~\ref{fig:APSK} shows the auto power spectrum of the convergence calculated using Eq.~(\ref{eq:kappa}) and the minimum variance reconstruction approximate noise spectrum for the Planck CMB observation from the temperature fluctuation and the polarization map. The signal is less than the noise at any $\it{\ell}$, and the noise is two orders of magnitude larger than the signal on small scales. Therefore, the error on the cross power spectrum on small scales is dominated by the noise of the convergence and makes the detection difficult even if the sensitivity of SKA-mid (IF) is sufficient to measure the 21cm line signal.

\begin{figure}
\includegraphics[width=\columnwidth]{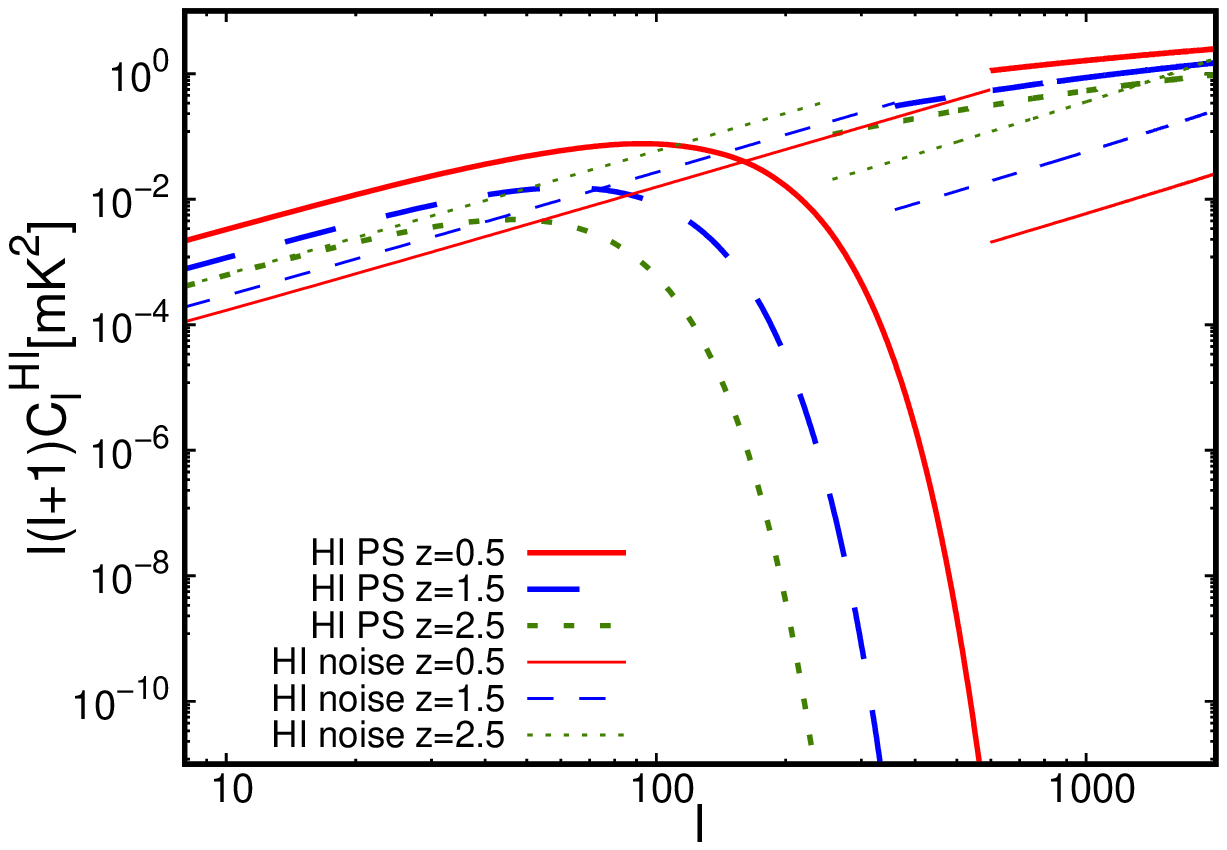}
\caption{
Sample variances (thick) and thermal noise of SKA1-mid observations per pointing (thin). The H$_{\rm I}$ signal is calculated using Eq.~(\ref{eq:21--cmAPS}). The gap of the noise curves indicates the change of observation mode (SD and IF).}
\label{fig:APSHI}
\end{figure}

Fig.~\ref{fig:APSHI} represents the H$_{\rm I}$ auto power spectrum and the thermal noise of the SKA-mid per pointing. The thermal noise is smaller than the expected signal on large scales ($\ell \lesssim 100$) at $z=0.5$ ,$1.5$ and on small scales ($\ell \gtrsim 600$) which are covered by the interferometer mode, and therefore a statistically-significant detection of the H$_{\rm I}$ signal using SKA1-mid is relatively easy in the absence of foregrounds. However, the noise is comparable to the signal at $z = 2.5$, and thus the thermal noise of 21\,cm line observation limits the detection of the cross power spectrum at large scales.

\begin{figure}
\includegraphics[width=\columnwidth]{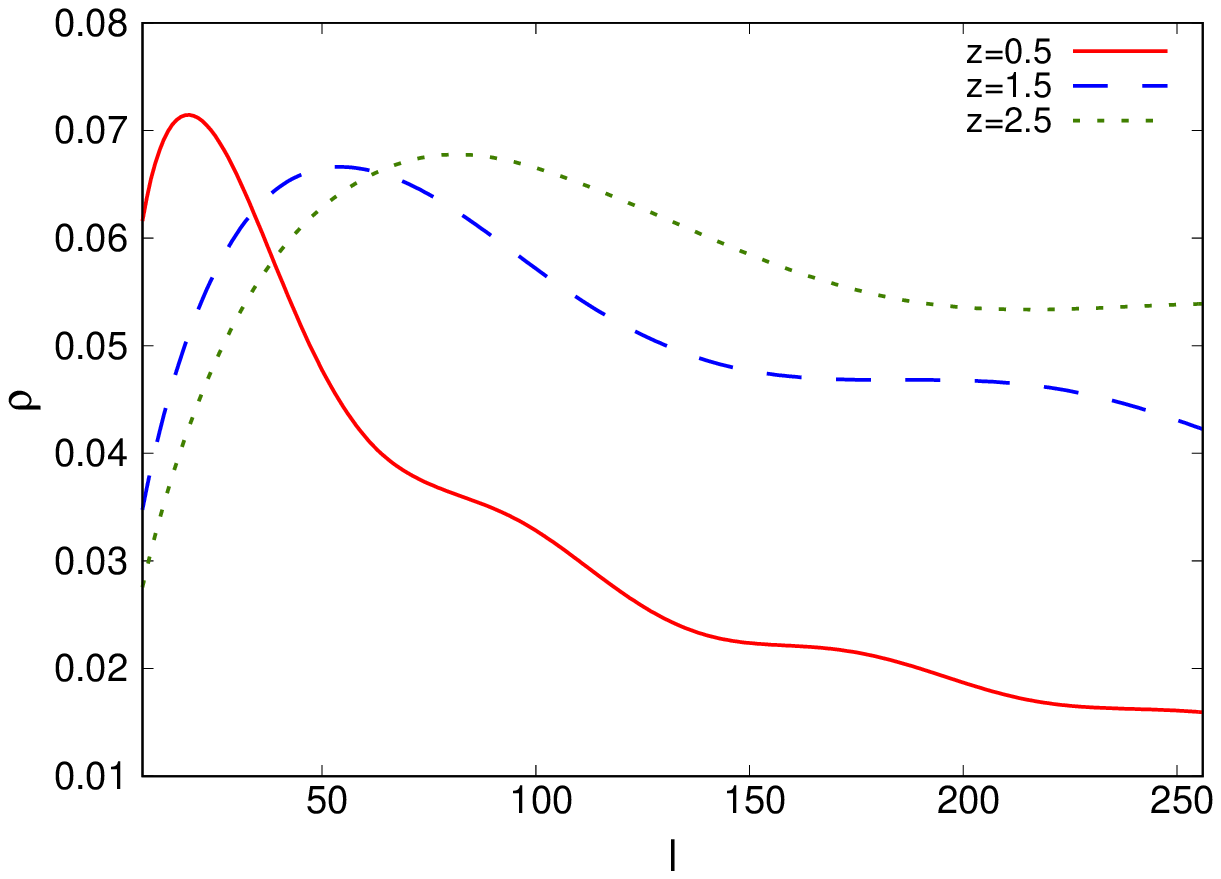}
\caption{Correlation coefficient between 21cm line and convergence calculated by Eq.~(\ref{eq:rho}).}
\label{fig:rho}
\end{figure}

The correlation coefficient, defined as 
\begin{equation}
\rho = \frac{C_{\ell}^{\rm H_{\rm I}-\kappa}}{\sqrt{C_{\ell}^{\kappa} C_{\ell}^{\rm H_{\rm I}}}},
\label{eq:rho}
\end{equation}
is shown in Fig.~\ref{fig:rho} and has a broad peak at $\ell \sim 20, 50$ and $80$ for $z = 0.5, 1.5$ and $2.5$, respectively, and is typically 0.05. This indicates that the factor $N_{c} f_{\rm sky}$ in Eq.~(\ref{eq:SN2}) must be larger than O(10) for a significant detection even if the instrumental noise terms ($N_{\ell}^{\kappa}$ and $N_{\ell}^{\rm H_{\rm I}}$) are negligible. 

\begin{figure}
\begin{flushleft}
\centering
\includegraphics[width=\columnwidth]{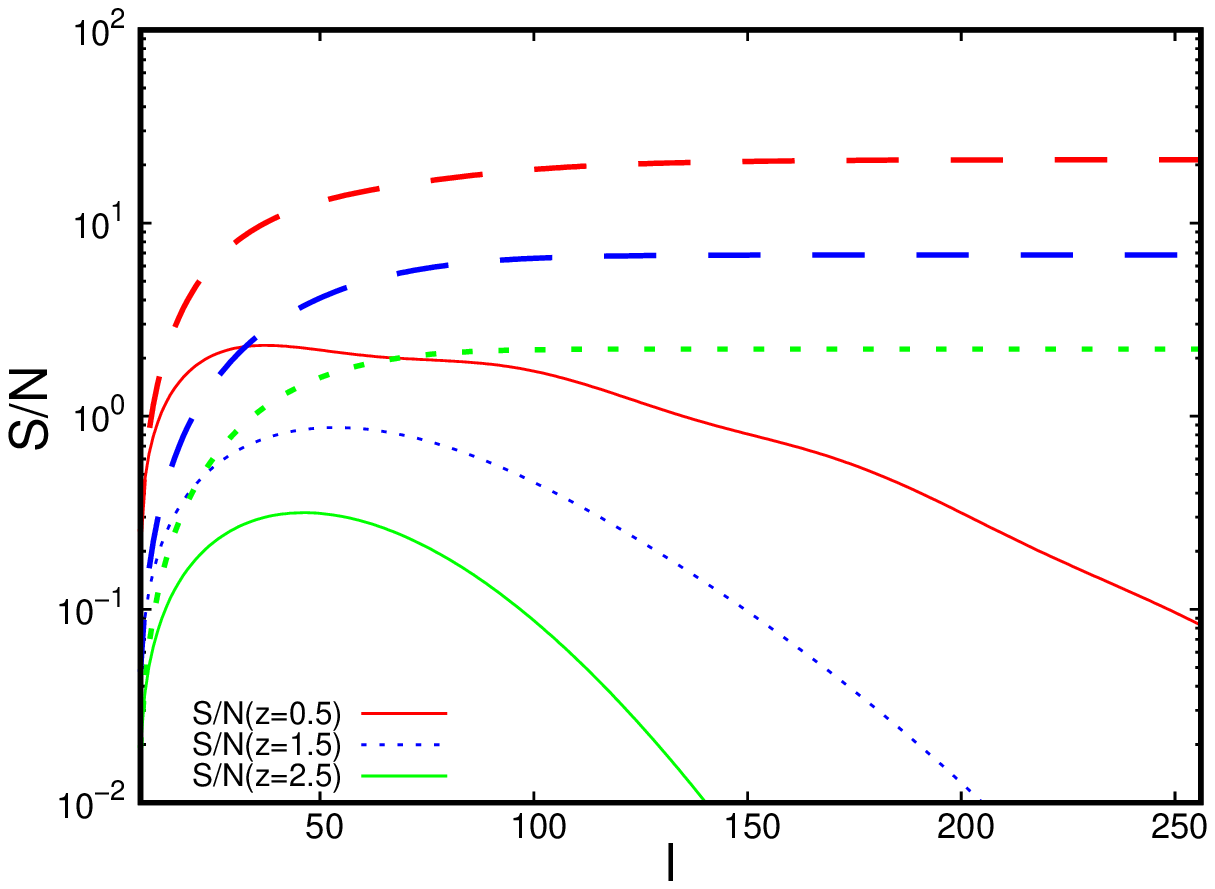}
\caption{Expected S/N ratio with large-scale 21cm-line observation with SKA1-mid SD mode. Thin lines represent S/N ratios for individual multipole modes, while thick lines represent cumulative ones.
}
\label{fig:SN}
\end{flushleft}
\end{figure}

Fig.~\ref{fig:SN} shows the S/N ratio for individual multipole mode and those accumulated over $\ell$, in a multipole range covered by the SKA-mid SD mode. The total S/N ratios in this range reach 20, 7 and 2 for $z = 0.5, 1.5$ and $2.5$, respectively. They are saturated at $l \sim 100$, and the damping of power due to the finite beam size have a small impact at $z = 0.5$ and $1.5$ and a large impact at $z = 2.5$. The higher multipole modes covered by SKA1-mid IF mode does not contribute to the cumulative S/N very much. In fact, the gain from the IF mode is only S/N $\sim$ 0.1 for the fiducial survey specification. This results indicates that the detection of $C_{\ell}^{\rm H_{\rm I}-\kappa}$ can be achieved with only at large scales $\ell \lesssim 200$ for the current survey parameters.

\begin{figure}
\centering
\includegraphics[width=8cm]{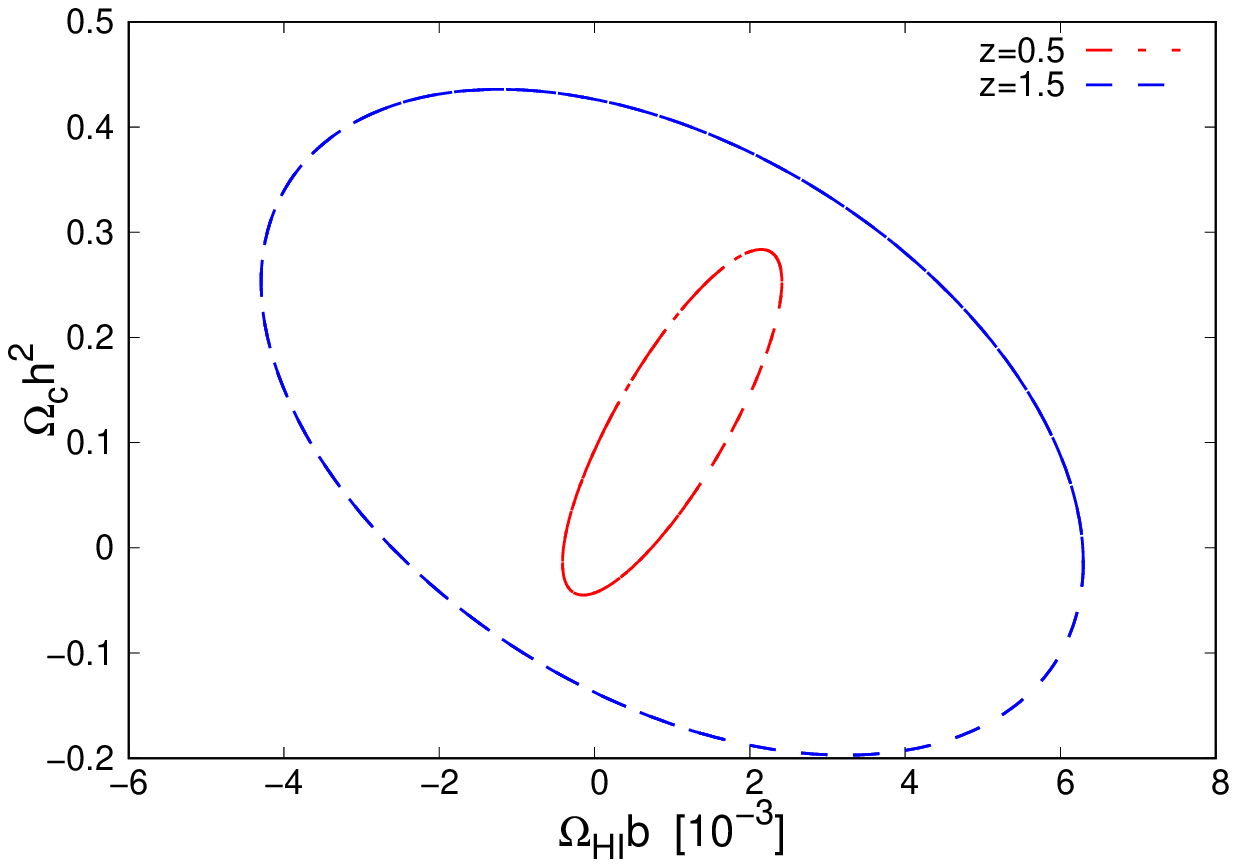}
\includegraphics[width=8cm]{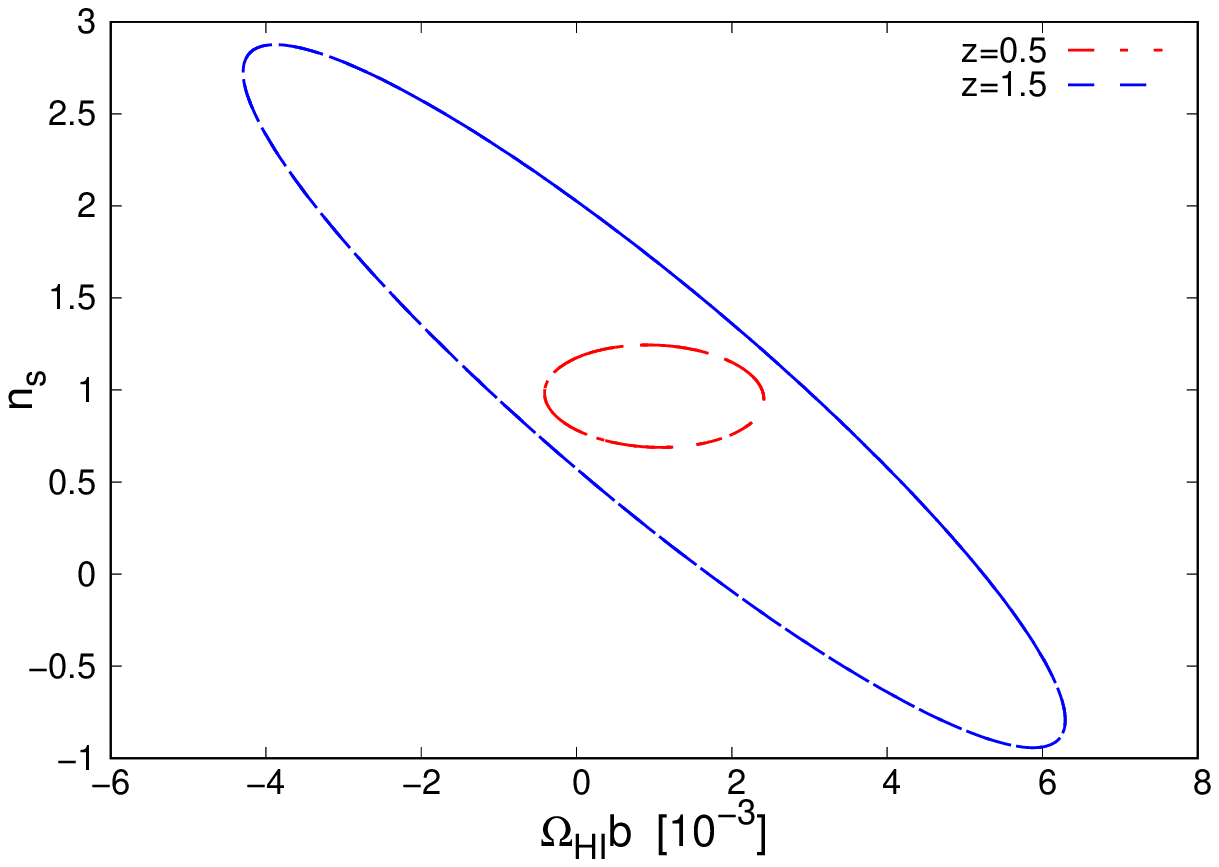}
\includegraphics[width=8cm]{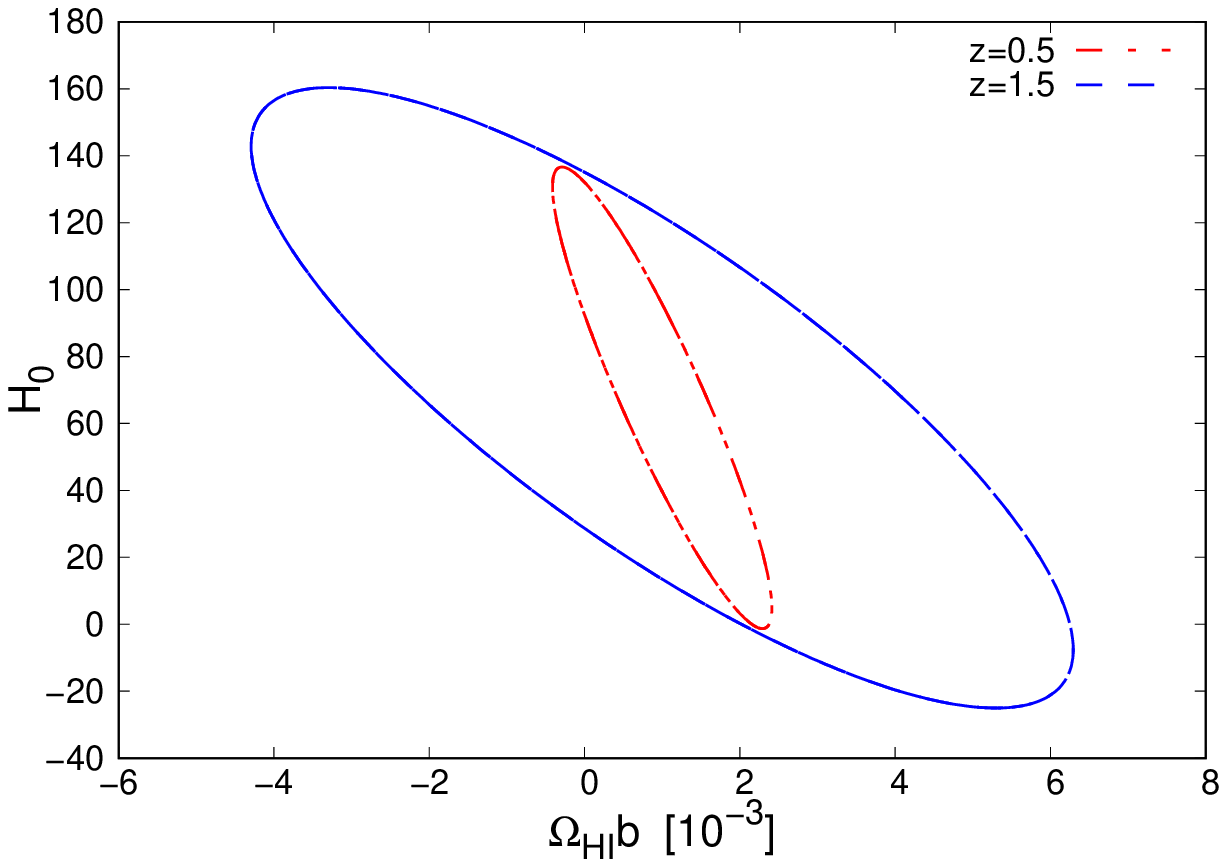}
\caption{Expected constraints on cosmological parameters from the CPS with large-scale 21cm-line observations by the SKA1-mid SD-mode. Two-dimensional constraints on ($\Omega_{\rm H_{\rm I}} b$, $\Omega_{\rm c} h^2$), ($\Omega_{\rm H_{\rm I}} b$, $n_{\rm s}$) and ($\Omega_{\rm H_{\rm I}} b$, $H_0$) planes are shown, where other parameters are marginalized.}
\label{fig:Fisher}
\end{figure}

Fig.~\ref{fig:Fisher} shows 68.5 $\%$ confidence regions on two parameters obtained by marginalizing over the other two parameters. The confidence regions at z=2.5 are so large that they are not shown in these Figures. If we compare the power of constraints among different redshifts, it is slightly stronger at lower redshifts than those for high redshift because lower redshifts have higher $\it{\rm S/N}$ ratio. In the case we do not use priors on the parameters from other observations, these constraints are rather weak and we cannot exclude $\OHI b = 0$.

\begin{figure}
\centering
\includegraphics[width=\columnwidth]{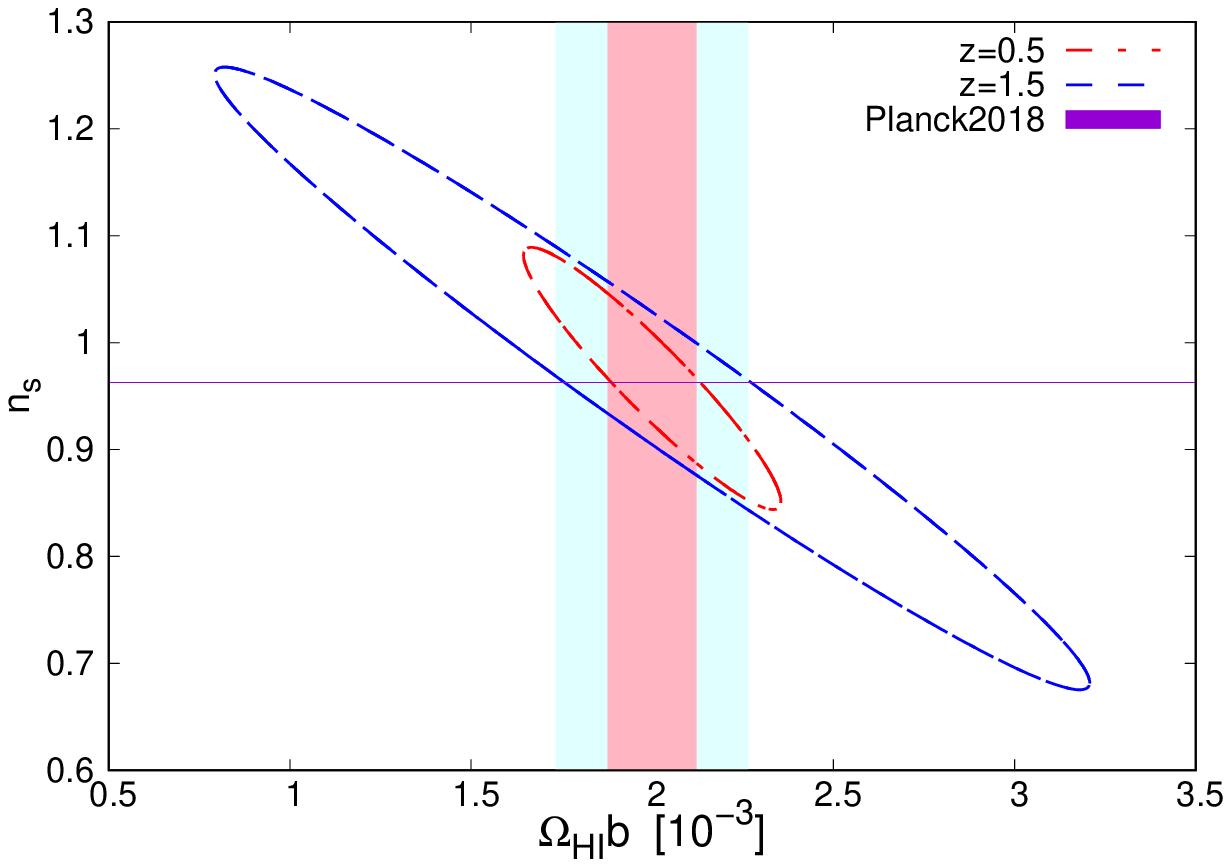}
\caption{Expected constraints on $\Omega_{\rm H_{\rm I}} b$ and $n_{\rm s}$ from the CPS with large-scale 21cm-line observations by the SKA1-mid SD-mode, fixing the other two cosmological parameters. The shaded regions, light blue for $z = 1.5$ and pink for $z = 0.5$, represent the constraints which can be obtained by fixing $n_{\rm s}$ to the value given by the Planck (horizontal line).}
\label{pic:HIns}
\end{figure}

Among the 4 parameters, the standard cosmological parameters, $\Omega_{\rm c} h^2, H_0$ and $n_s$, have already been determined precisely by other observations with typical errors less than 1 \% and we can use them as priors. Thus, we demonstrate the combined constraints on $\Omega_{\rm H_{\rm I}} b$ by simply fixing the other cosmological parameters, that is, removing them from Fisher matrix. This treatment gives a reasonable estimate of an expected constraint because the precision of the three cosmological parameters is much better than expected from the CPS. In Fig.~\ref{pic:HIns}, we show expected constraints on $\Omega_{\rm H_{\rm I}} b$ and $n_s$ obtained by fixing $\Omega_{\rm c}h^2$ and $H_0$ to the Planck values. The constraints improve remarkably and the expected error on $\Omega_{\rm H_{\rm I}} b$ is about 17.8 \% for $z=0.5$ and 60 \% for $z=1.5$. However, we cannot exclude $\OHI b = 0$ only at $z=2.5$ (not shown).

Further, fixing $n_s$ as well, the expected constraints on $\Omega_{\rm H_{\rm I}} b$ become even better and are typically $6-13~\%$. In order to obtain constraints on $\Omega_{\rm H_{\rm I}}$ itself, we need other independent observations such as H$_{\rm I}$ auto-correlation to measure the H$_{\rm I}$ bias. If the H$_{\rm I}$ bias is measured with a similar precision, the resultant constraints on $\Omega_{\rm H_{\rm I}}$ will also be of order $6-13~\%$, which are better than previous constraints \citep{Lah,Khandai,Rao}.

The detection of the CPS at small scales with the SKA-mid IF mode is not likely because the cumulative S/N ratio reaches only $O(0.1)$. This is due to the sensitivity of the Planck at small scales. Other instruments with a high angular resolution such as ACTPol \citep{niemack,thornton} will reduce significantly the noise at small scales and allows us to constrain cosmological parameters even more precisely. In addition, recently, Subaru Hyper Supreme-Cam has made a wide field optical weak lensing map \citep{Oguri2018}. Although the measured area is small, the data should be suitable for the cross correlation with the IF mode observation at low redshifts.

\begin{figure}
\includegraphics[width=\columnwidth]{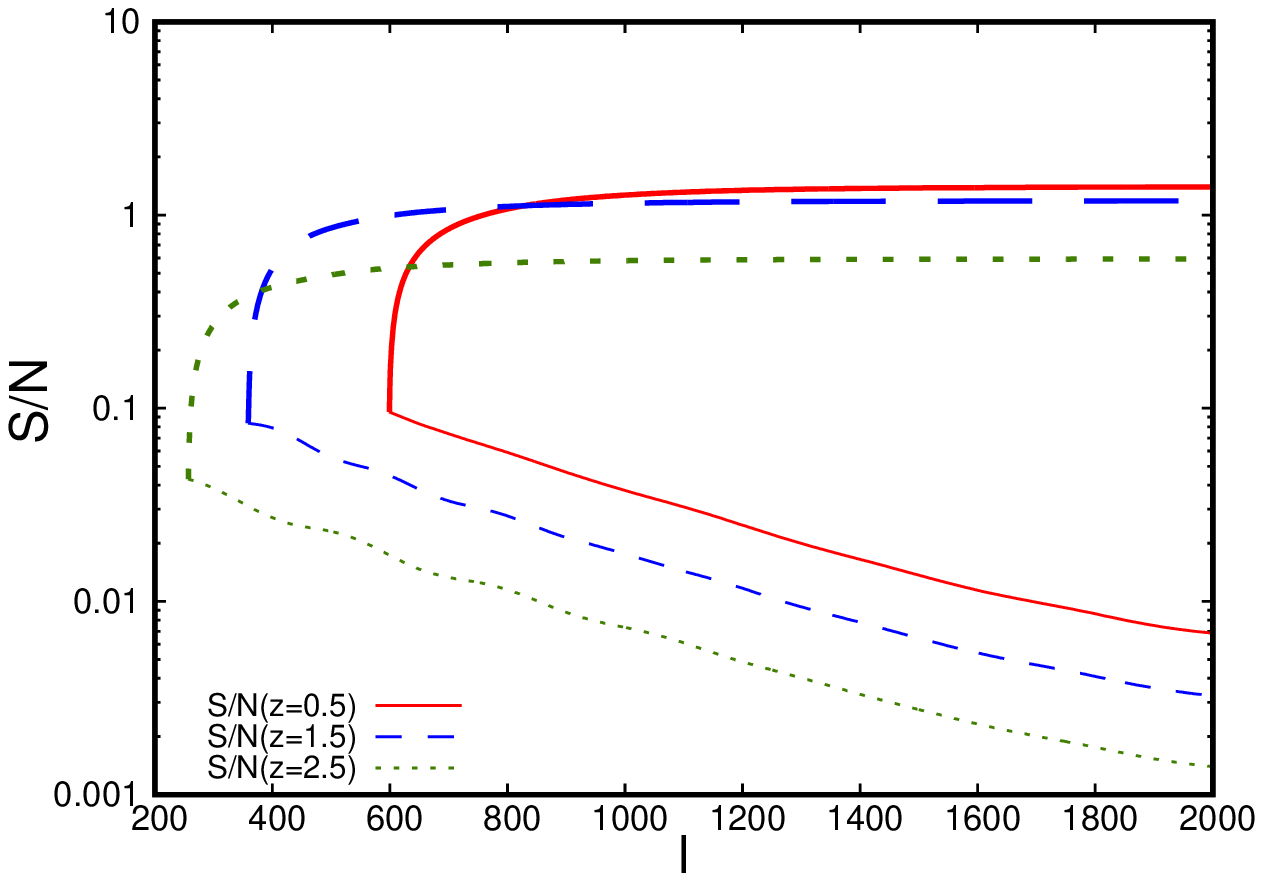}
\caption{Expected S/N ratio with small-scale 21cm-line observation with SKA1-mid IF mode. Thin lines represent S/N ratios for individual multipole modes, while thick lines represent cumulative ones. Here the observation time per pointing is set to 1 hour, while the total observation time is fixed to 1,000 hours.}
\label{fig:fsky1000}
\end{figure}

As we can see in Fig. \ref{fig:APSHI}, the sample variance dominates the thermal noise in the CPS error, especially at small scales covered by IF mode. Therefore, with a fixed total observation time, we can further reduce the error effectively by expanding the survey area and reducing observation time per pointing. Although an investigation of the optimal survey area is beyond the scope of the current paper, we demonstrate an example of IF-mode observation in Fig. \ref{fig:fsky1000}. Here, the survey area is widened by a factor of $10^3$, while the observation time per pointing is reduced by a factor of $1/10^3$. We can see that the total S/N of small-scale CPS reaches $O(1)$.

For practical observations of the 21cm-line signal, the Galactic and extragalactic emissions are serious obstacles. There have been many studies to remove or avoid these foregrounds, and one of the common methods is to utilize the spectral smoothness of the foreground emission. However, if the smooth components are removed from the observed spectrum of each line of sight, the cross-correlation with the CMB lensing is expected to be reduced, because the CMB lensing signal is a projection from the last scattering surface and the observer and correlates mostly with the large-scale fluctuations in the 21cm-line signal. On the other hand, the foregrounds could be removed or avoided also by using their spatial smoothness and pattern. In this case, the cross-correlation with the CMB lensing would not be degraded significantly. Because the purpose of the current paper is to evaluate the intrinsic sensitivity of the cross-correlation on the determination of cosmological parameters, we hope to study this important problem elsewhere.

In addition to the observational challenges, there is an uncertainty in the bias model between the cold dark matter and H$_{\rm I}$ gas fluctuations. In this work, we assumed a constant linear bias model, which is appropriate on large scales. The non-linear scale- and redshift-dependent bias model has been studied using numerical simulation in the literature \citep{2016MNRAS.460.4310S,VN2018,Ando2019}. 

\section{Summary}
\label{sec:summary}
In this work, we have studied the detectability of the 21\,cm line-CMB lensing cross power spectrum using a practical model of instrumental noise. For the observation of the 21\,cm signal, we assumed the SKA1-mid operating as single-dish mode and interferometer mode and considered 1,000-hour observations of Band 1 and Band 2 for each mode, resulting in the total observation time of 4,000 hours. For the modeling of the noise of convergence map, we referred to the recent Planck result. We found that the cross power spectrum can be detected at large scales ($\ell \lesssim 200$) with a total S/N larger than 7 for a redshift range of 0-2 by combining the Planck and SKA1-mid single-dish mode. However, we do not expect a significant detection on small scale ($\ell \gtrsim 300$) with SKA1-mid interferometer mode due to the limited angular resolution of the convergence field. Small-scale-oriented instruments such as ACTPol will be able to reduce the noise on the convergence map, and thus we expect more significant detection of the cross power spectrum on small scales.

Next, we performed Fisher analysis to estimate expected constraints on cosmological parameters, $\Omega_{\rm c}h^{2}, n_{s}$, $H_0$ and $\Omega_{\rm H_{\rm I}} b$, considering that $\Omega_{\rm H_{\rm I}}$ always appears in the cross power spectrum with the bias $b$. Although the constraints from the cross power spectrum alone are rather weak, by fixing $\Omega_{\rm c} h^{2}, n_{s}$ and $H_0$ which have been already well constrained by the Planck, $\Omega_{\rm H_{\rm I}} b$ can be constrained with a precision of 6-13$\%$ at $z = 0.5$ and $1.5$. These results are based on the constant linear bias model between cold dark matter and H$_{\rm I}$ gas fluctuations, which is appropriate on the large scale. If the bias is constrained from other observations, $\Omega_{\rm H_{\rm I}}$ can be directly constrained.

The total observation time assumed in this paper is 4,000 hours, considering SKA1-mid Band 1 and 2 observations with both SD and IF modes. Although this looks a huge number, because IF mode is not effective and Band 2 contributes only to a half of one of three redshift bins, most of the information come from Band 1 observation with SD mode. Thus, 1,000 hours of SD-mode Band 1 observation are sufficient for most of our purpose.

\acknowledgments

This work is supported by JSPS KAKENHI Grant Numbers,
JP16J01585, JP26610048, JP15H05896, JP16H05999, JP17H01110, JP15H05890, and Bilateral Joint Research Projects of JSPS. KT's work is partially supported by the ISM Cooperative Research Program 2020-ISMCRP-2017.


\end{document}